\documentclass{aa}
\usepackage{graphicx,natbib}
\bibpunct{(}{)}{;}{a}{}{,}

\def\lum{\rm erg~ s$^{-1}$}

\begin{document}

\sloppypar

   \title{Properties of the Galactic population of cataclysmic variables in
   hard X-rays}

   \author{M.~Revnivtsev\inst{1,2}, S.~Sazonov \inst{1,2}, R.~Krivonos \inst{2,1}, H.~Ritter\inst{1}, R.~Sunyaev \inst{1,2}}

   \offprints{mikej@mpa-garching.mpg.de}

   \institute{
              Max-Planck-Institute f\"ur Astrophysik,
              Karl-Schwarzschild-Str. 1, D-85740 Garching bei M\"unchen,
              Germany,
     \and
              Space Research Institute, Russian Academy of Sciences,
              Profsoyuznaya 84/32, 117997 Moscow, Russia
            }
  \date{}

        \authorrunning{ M.~Revnivtsev et al.}

\abstract{We measure the spatial distribution and hard X-ray
luminosity function of cataclysmic variables (CVs) using the
INTEGRAL all-sky survey in the 17--60 keV energy band. The vast
majority of the INTEGRAL detected CVs are intermediate polars with
luminosities in the range $10^{32}$--$10^{34}$~erg~s$^{-1}$. The scale
height of the Galactic disk population of CVs is found to be
$130^{+90}_{-50}$~pc. The CV luminosity function   
measured with INTEGRAL in hard X-rays is compatible with that
previously determined at lower energies (3--20~keV) using a largely
independent sample of sources detected by RXTE (located at $|b|>10^\circ$ as 
opposed to the INTEGRAL sample, strongly concentrated to the Galactic
plane). The cumulative 17--60~keV luminosity density of CVs per unit
stellar mass is found to be $(1.3\pm0.3)\times10^{27}$ \lum\ 
$M^{-1}_\odot$ and is thus comparable to that of low-mass X-ray
binaries in this energy band. Therefore, faint but numerous CVs are expected to provide an
important contribution to the cumulative hard X-ray emission of galaxies.   
}
 
   \maketitle

%

\section{Introduction}
The sky is much less studied in hard X-rays ($>$20--50~keV) than at
lower energies because of difficulties with 
detecting hard X-ray photons and, more importantly, due to the
rapidly decreasing photon flux from cosmic sources with increasing
energy. The largest existing catalogs comprise only about $\sim$400
hard X-ray sources \citep{krivonos07b,bird07,tueller07}. As 
a result, basic properties of different populations of hard X-ray
sources, such as number density, luminosity function, spatial distribution,
etc., are poorly known. Therefore, any sample of hard X-ray sources
with well-defined statistical properties (such as that of 
\citealt{krivonos07b}) is of high value, since it allows one to
study the parameters of source populations. 

The scarceness of information about the Galactic populations of hard X-ray
sources made it difficult to interprete the results of
some hard X-ray observations. For example, a number of early
experiments detected 
unresolved hard X-ray emission from the Galaxy and in particular from 
its central region, which was suggested to be truly diffuse
emission arising from interaction of low-energy cosmic rays with
the interstellar medium
\citep[e.g.][]{stecker77,mandrou80,sacher84,skibo93}. However, the 
INTEGRAL observatory \citep{integral} has recently resolved most of
the Galactic hard X-ray emission at energies below $\sim 100$~keV into
a number of 
luminous discrete sources \citep{lebrun04}. Furthemore, the remaning unresolved
emission -- the so-called Galactic ridge X-ray emission (GRXE) -- has been
shown to closely trace the stellar mass distribution in the Galaxy,
suggesting that it might be composed of a large number of fainter
stellar-type sources \citep{krivonos07a}. 

For the stellar origin of the GXRE to be correct, the space density of
faint Galactic hard X-ray sources must be sufficiently high to provide the
required luminosity density. At energies above several keV, the
main contribution to the GRXE is expected from cataclysmic variables
(CVs), and more specifically from intermediate polars (IPs,
\citealt{sazonov06,revnivtsev06}). Using the luminosity function of
CVs measured in the 3--20~keV band \citep{sazonov06} 
and assuming some fiducial shape of the spectrum of a generic
intermediate polar, \cite{revnivtsev06} and \cite{krivonos07a}
demonstrated that such sources can indeed produce all of the unresolved
Galactic hard X-ray emission. Nonetheless, although the spectra
of intermediate polars measured individually with INTEGRAL \cite[see
e.g.][]{revnivtsev04a,revnivtsev04b,barlow06} in general agree 
with the spectrum assumed by \cite{krivonos07a}, it is still important
to directly measure the cumulative luminosity density of CVs at energies higher
than 20~keV.

In order to obtain direct estimates of the luminosity function 
and cumulative luminosity density of faint Galactic hard X-ray sources, we 
have analized results of the INTEGRAL all-sky survey \citep{krivonos07b}.
Apart from low and high mass X-ray binaries, the INTEGRAL catalog contains
Galactic sources of other types, mostly cataclysmic variables. The only
non-CV faint Galactic sources in the INTEGRAL catalog are the
symbiotic stars RT Cru and IGR J10109$-$5746/CD-57 3057 and the peculiar
stars Gamma Cas and Eta Carinae. Because of the very small number of such
sources, we do not consider them here and focus on studying the
properies of CVs detected in hard X-rays.

\section{The sample of sources}

We selected all INTEGRAL sources from the catalog of
\cite{krivonos07b} identified as cataclysmic variables.
This sample ideally suits our puproses because it is based
on an all-sky survey and serendipitous with respect
to CVs (no deep INTEGRAL exposures were devoted to observations of 
CVs from the sample). We selected only those sources detected with
more than $5\sigma$ significance on the all-time averaged map of the
sky. The resulting sample, consisting of 15 intermediate polars and 1 dwarf
nova, is presented in Table~\ref{table}.

\begin{figure}[htb]
\vbox{
\includegraphics[width=0.8\columnwidth,bb=34 450 571 700,clip]{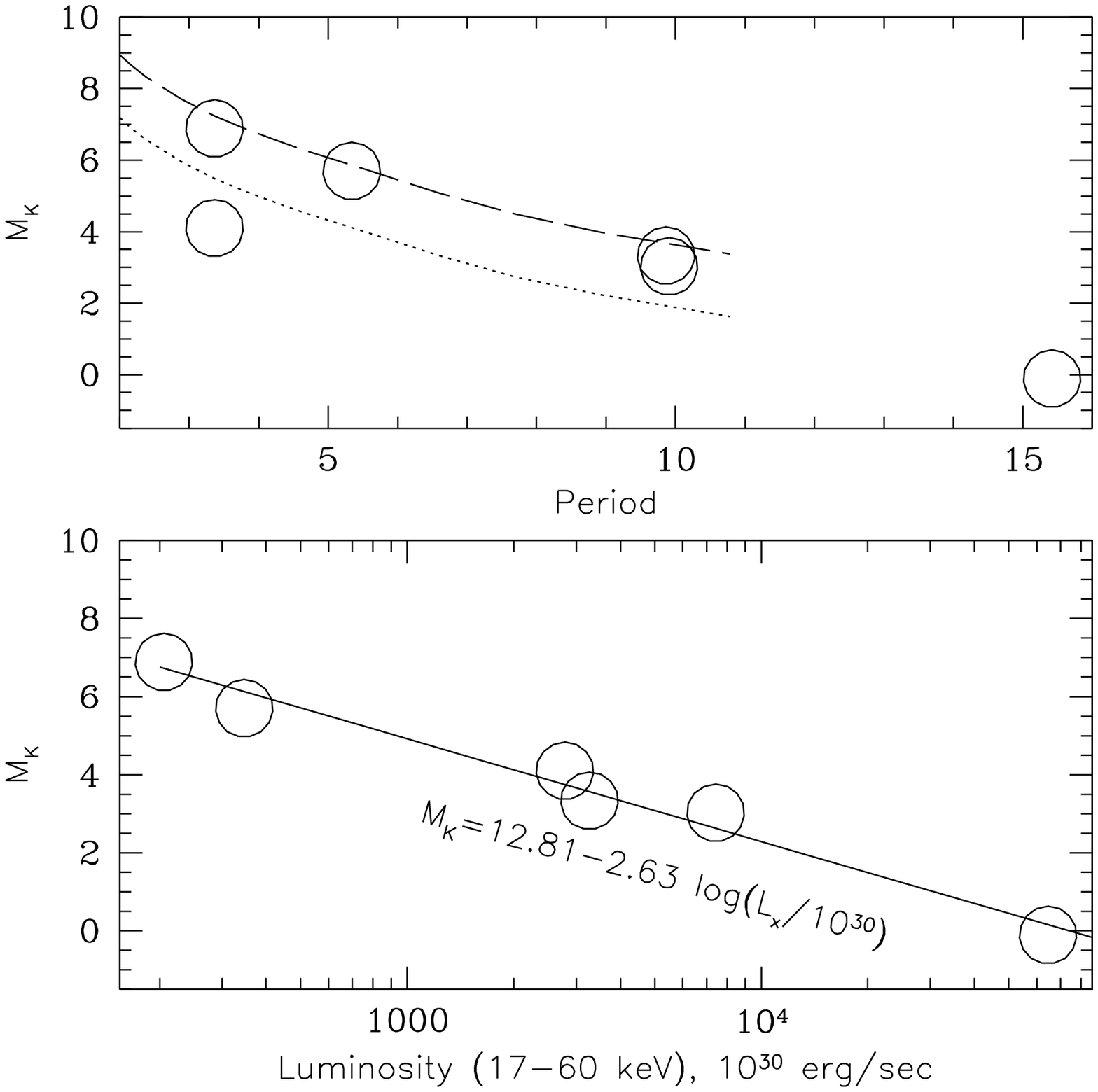}
\includegraphics[width=0.8\columnwidth,bb=34 180 571 450,clip]{./correlations_known_dist.ps}
\includegraphics[width=0.8\columnwidth,bb=34 184 567 701,clip]{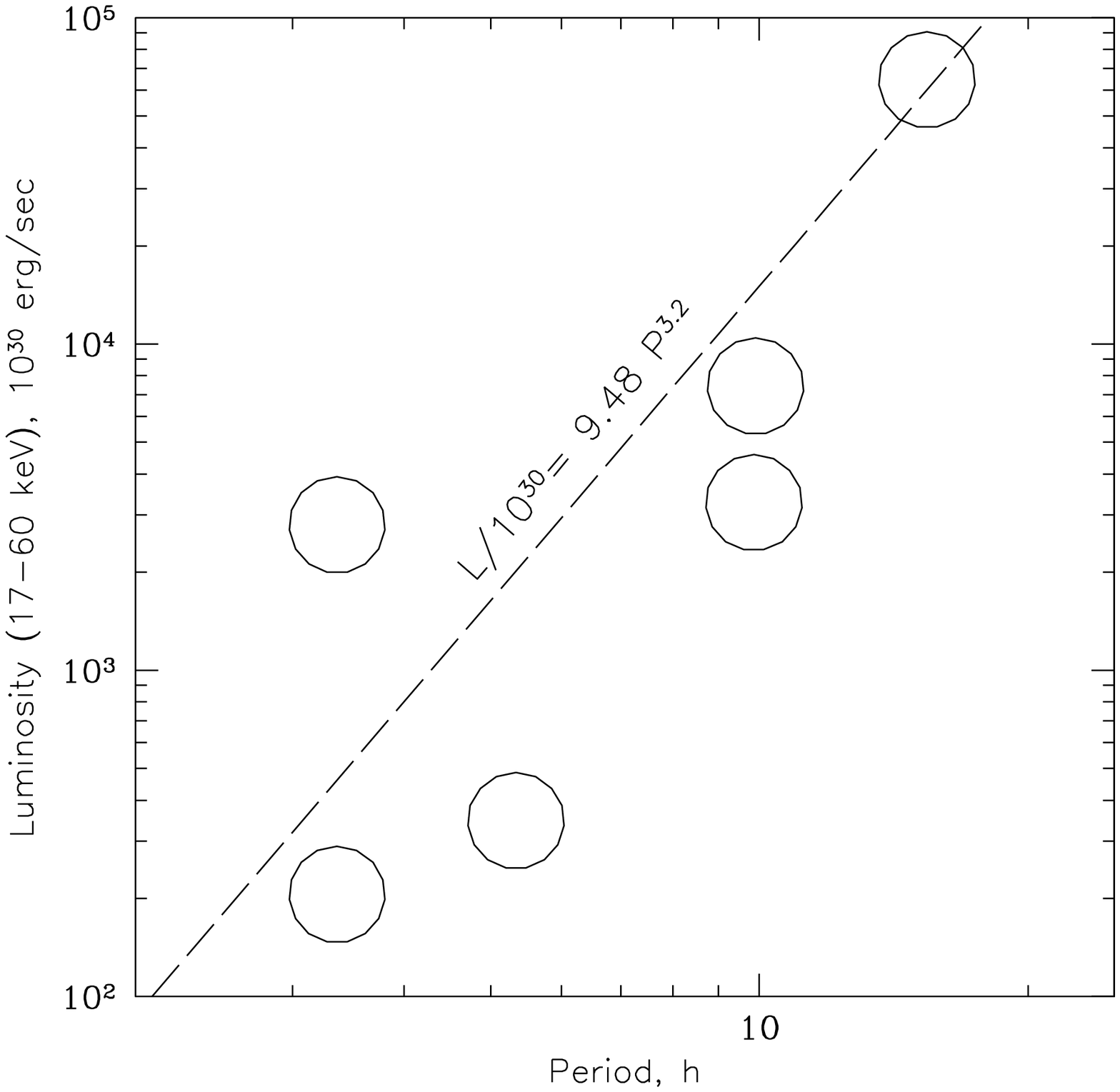}
}
\caption{ {\sl Upper panel} -- dependence of the absolute $K$-band
brightness of IPs on their orbital period. Circles are measurements for
sources from our sample. The dashed line shows the theoretical estimate 
of $M_K$ based on \cite{baraffe98}. The dotted line is the same
estimate shifted by 1.75 mag, which is equivalent to assuming that the
secondary star contributes only 1/5 to the total near-infrared brightness of an
IP. {\sl Middle panel} -- absolute $K$-band brightness of IPs as a
function of their hard X-ray luminosity. An empirical fit is shown by
the solid line. 
{\sl Lower panel} -- correlation of the hard X-ray (17--60~keV)
luminosity of IPs with their orbital period. The dashed line shows the
empirical dependence $L_{\rm x}\propto P^{3.2}$.  
}
\label{corr}
\end{figure}

\subsection{Source distances}

Accurate distance estimates are not available for more than half of
the sources in our sample, which  presents a serious obstacle
for constructing a CV luminosity function.

Usually, if the spectral type of the secondary star and the apparent brightness
of a binary system are known, one can try to estimate its distance
\cite[e.g.][]{barnes74,beuermann06,knigge06}. However, in the case of intermediate
polars (which constitute the majority of our sample), the accuracy of this
method is significantly affected by the fact that the optical light of an
IP is produced not only by the secondary star \cite[see e.g.][]{knigge06}, but also by the
accretion disk \cite[see e.g.][]{beuermann04} and white dwarf surface
\cite[e.g.][]{bonnet01}. Detailed spectroscopic information
allows one in some cases to determine the contribution of the light of
the companion star to the optical brightness of the binary
\cite[e.g.][]{watson95,bonnet01,gansicke05}, but such information is
not available for all CVs in our sample.

\begin{table*}
\caption{CVs from the INTEGRAL all-sky survey
\citep{krivonos07b}. Distance estimates enclosed in parentheses were
estimated either assuming an absolute $K$-band brightness of the
binary system or using the period--luminosity relation.}
\begin{tabular}{lrrrrrcccl}
Name    &$l^{II}$&$b^{II}$& Type &Orb.per.(h)&$F_{\rm 17-60~keV}$& D,pc&$m_K$& $\log L_{\rm hx}$ & $1/V_{\rm gen}$\\ 
\hline
V709 Cas         &120.04& -3.45&IP&5.34&3.91& 230$^1$ &12.51& 32.54 &$1.09\times10^{-8}$\\
GK Per           &150.95&-10.10&IP&47.9&1.82& 470$^2$ &10.06& 32.80 &$5.08\times10^{-9}$\\
V1062 Tau        &178.08&-10.31&IP&9.90&3.67&1100$^3$ &13.24& 33.87 &$2.86\times10^{-10}$\\
IGR J14536-5522  &319.74&  3.46&IP&    &1.20&(347)$^4$&12.70& 32.38 &$1.74\times10^{-8}$\\
IGR J15094-6649  &315.92& -7.49&IP&    &1.05&(496)$^4$&13.48& 32.64 &$8.20\times10^{-9}$\\
NY Lup           &332.44&  7.02&IP&9.86&4.10& 690$^5$ &12.52& 33.51 &$7.08\times10^{-10}$\\
IGR J16167-4957  &349.70&  7.33&IP&    &1.45&(546)$^4$&13.76& 32.86 &$4.27\times10^{-9}$\\
IGR J16500-3307  &359.87&  8.74&IP&    &1.12&(552)$^4$&13.71& 32.76 &$5.75\times10^{-9}$\\
V2400 Oph        &359.86&  8.74&IP&3.42&2.64&(331)$^6$&13.02& 32.55 &$7.10\times10^{-9}$\\
IGR J17195-4100  &346.98& -2.11&IP&    &1.92&(217)$^4$&11.69& 32.18 &$3.26\times10^{-8}$\\
IGR J17303-0601  & 17.90& 15.01&IP&15.4&3.54&3300$^7$ &12.48& 34.81 &$3.47\times10^{-11}$\\
4U 1849-31       &  4.96&-14.35&IP&3.36&6.40& 510$^8$ &12.64& 33.44 &$8.49\times10^{-10}$\\
RX J1940.2-1025  & 28.98&-15.50&IP&3.36&2.32&230$^9$  &13.69& 32.31 &$2.16\times10^{-8}$\\
IGR J21237+4218  & 87.11& -5.68&IP&7.48&1.03&(1854)$^7$&13.73&33.58 &$3.65\times10^{-10}$\\
IGR J21335+5105  & 94.36& -0.40&IP&7.19&3.15& (995)$^7$&13.45&33.88 &$4.20\times10^{-10}$\\
SS Cyg           & 90.56& -7.11&DN&6.60&2.89&166$^{10}$& 8.29&32.12 &$3.87\times10^{-8}$\\
\hline
\end{tabular}
\begin{list}{}
\item  (1)- \cite{bonnet01}, (2)- \cite{slavin95},
(3)-\cite{patterson94}, (4)- distance was calculated assuming $K$-band
absolute magnitude $M_K=5$, (5) - \cite{demartino06}, (6) based on
period--hard X-ray luminosity correlation, (7) - \cite{sazonov06},
based on \cite{gansicke05}, (8) - \cite{beuermann04},
(9)-\cite{watson95}, (10) - \cite{bitner07} 
\end{list}
\label{table}
\end{table*}

\begin{table*}
\label{srcs_dist}
\caption{Sources from the main sample with available information on distances.}
\begin{center}
\begin{tabular}{lrcrrrr}
Name             &D,pc        &Orb.~period, h& $F_{\rm hx}$&   $\log L_{\rm hx}$    &  $m_K$     &   $M_K$    \\
                 &            &         &   mCrab     &                       &            & \\
\hline
V709 Cas         &    230$^1$ &5.34     &3.91 &      32.54 &      12.51 &       5.70 \\ 
V1062 Tau        &   1100$^2$ &9.90     &3.67 &      33.87 &      13.24 &       3.03 \\ 
NY Lup           &    690$^3$ &9.86     &4.10 &      33.51 &      12.53 &       3.33 \\ 
IGR J17303-0601  &   3300$^4$ &15.4     &3.54 &      34.81 &      12.49 &      -0.11 \\ 
4U 1849-31       &    510$^5$ &3.36     &6.40 &      33.45 &      12.64 &       4.10 \\ 
RX J1940.2-1025  &    230$^6$ &3.36     &2.32 &      32.31 &      13.70 &       6.89 \\ 
\hline
\end{tabular}
\end{center}
\begin{list}{}
\item (1) -- \cite{bonnet01}, (2) -- \cite{patterson94}, (3) -- \cite{demartino06}, (4) -- \cite{sazonov06}, based on \cite{gansicke05}, (5) -- \cite{beuermann04}, (6) -- \cite{watson95}
\end{list}
\end{table*}

In order to demonstrate the importance of the contribution of the accretion 
disk to the NIR emission of our IPs, we selected those sources with
available distance estimates (Table \ref{srcs_dist}). In all these
6 systems, the companion star belongs to the main sequence (we excluded
here the IP with a giant companion GK Per and dwarf nova SS
Cyg due to unclear correlation of its measured NIR brightness
with the mass tranfer rate in the binary system, set by the orbital period) 
and therefore their absolute brightness can be
related to their size. 

The orbital period ($P$) of such systems can
be used to estimate the radius of the secondary star ($R_2$) under the
assumption that the star fills its Roche lobe:
$$
R_2/R_\odot\approx 0.234 (M_2/M_\odot)^{1/3} P^{2/3}.
$$
Assuming the mass--radius relation $R \approx 0.91 M^{0.75}$ for the
late-type main sequence stars in CVs \citep[e.g.][]{warner95,smith98},
we can write $R_2/R_\odot\approx0.081 P^{1.18}$ \citep{smith98}, where
the orbital period $P$ is measured in hours. In Fig.~\ref{corr} we
compare the expected dependence (using \citealt{baraffe98}) of
the absolute $K$-band brightness of the secondary star on binary
system period with the measured values for our CVs with known
distances. We see that while for some sources the observed and
predicted $K$-band absolute magnitudes agree within one magnitude,
other sources are significantly brighter in the near-infrared than
expected. 

This deviation results from the unaccounted contribution of
emission from the accretion disk in the binary system. 
Indeed, our sources demonstrate a good correlation of absolute
$K$-band brightness with hard X-ray luminosity (see Fig.~\ref{corr},
middle panel). The linear fit to the observed correlation gives:
$$
M_K\approx12.81-2.63\log L_{\rm hx,30},
$$
where $L_{\rm hx,30}$ is the source hard X-ray (17--60~keV) 
luminosity\footnote{Due to the relatively narrow range of spectral
shapes of IPs in the hard X-ray energy band \cite[see
e.g.][]{suleimanov05}, the hard X-ray luminosity might be considered a
proxy of the source bolometric luminosity or mass accretion rate.}
in units of $10^{30}$~erg~s$^{-1}$ .

This correlation is partly caused by the direct contribution of accretion 
disk emission to the optical and near-infrared light of the binary. Another 
contributor to this correlation is the anticipated dependence of the
mass accretion rate in IPs on the orbital period due to enhanced
magnetic stellar wind braking in binaries with larger optical stars
\citep{skumanich72,rappaport83,patterson84,postnov05}.  
Therefore, one can significantly underestimate the absolute
near-infrared brightness and distance of an IP if no allownace is
made for the X-ray luminosity.

On the other hand, we can use the $P$--$L_{\rm hx}$
correlation calibrated on our 6 CVs with known distances to predict 
hard X-ray luminosities and distances for other CVs from 
the INTEGRAL sample. The power-law slope of the empirical fit to the
observed luminosity (mass-accretion rate)--orbital period correlation
(Fig.~\ref{corr}, lower panel), 
$$
L_{\rm hx,30}\approx9.48 P^{3.2},
$$
agrees with previous empirical \cite[see e.g.][]{patterson84} and
theoretical estimates \cite[e.g.][]{postnov05}, indicating that this
dependence is physically motivated. 

We used the above empirical correlation to estimate the distances
to those sources in our sample with measured orbital periods. For the
remaning systems we made a rough estimate of their distance assuming
$M_K=5$. Although this assumption is certainly inaccurate (though
reasonable), it should not dramatically affect our analysis. For
example, if instead of $M_K=5$ we assumed $M_K=3$ or $M_K=7$, the
resulting value of the cumulative luminosity density of CVs would
change by less than a factor of 1.5. Nevertheless, one should keep in
mind that since the distances are not known with adequate accuracy for
all sources in the sample, all the properties of the Galactic CV population
estimated below contain systematic uncertainties in
addition to pure statistical ones.

\section{Spatial distribution}

Cataclysmic variables are not uniformly distributed around the Sun. 
Being relatively old stellar systems, they are concentrated
towards the Galactic plane: 
$\rho_{\rm CV}\propto \exp(-z/h)$, where $z$ is the height above the 
Galactic plane. The typical scale height $h$ of the distribution of
late-type stars in the Galaxy is 120--150~pc. Existing estimates for
CVs fall into this range \cite[e.g.][]{patterson84,ak08}. We do not consider
here the Galactocentric dependence of the CV distribution, since it
only becomes important at distances larger than 1.5--2~kpc from the
Sun and all of our sources except for one are located closer (we exclude this
source, IGR~J17303$-$0601, from the following determination of the
scale height of the CV spatial density distribution).

\begin{figure}[htb]
\centerline{
\includegraphics[width=0.8\columnwidth,bb=54 180 570 700,clip]{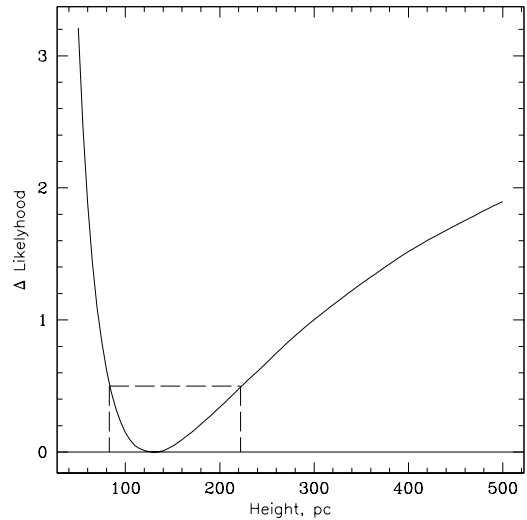}
}
\caption{Value of the maximum likelihood as a function of an assumed
exponential scale height $h$ of the CV spatial density
distribution. The 68\% confidence interval is indicated be the dashed lines.
}
\label{height}
\end{figure}

Using our INTEGRAL sample of sources derived from an all-sky survey we
can make our own estimate of the scale height of the Galactic disk
population of CVs. To this end, we constructed a maximum likelihood
estimator of the scale height $h$. The probability for a source to lie
at distance $z$ from the Galactic plane depends on $h$:
$P(z)dz=(1/h) \exp(-z/h)dz$. For each CV from the sample, we
calculated the maximum distance $D(\vec{\Omega})$ in a given direction
$\vec{\Omega}$ to which the source is detectable by the survey. The
shape of the resulting volume $V$ sampled by the survey over the whole
sky was then taken into account in calculating the probability of
source $i$ to lie at height $z$: $P_i^{V}(z)dz$. The sum of the
logarithms of such probabililities for all the sources in the sample
(except for IGR J17303$-$0601) defines our maximum
likelihood estimator: $ML=-\sum{\ln P_i^{V}}$. 

The dependence of the $ML$ value on $h$ is shown
in Fig.~\ref{height}. The best-fit value of the scale height is
130~pc, with the 68\% confidence interval of 84--223~pc. This agrees
with previous estimates of the CV scale height
\cite[e.g.][]{patterson84,ak08}, suggesting that most of our source
distance estimates are reasonable.

Since the $h$ value estimated above has a relatively large
uncertainty, we adopt $h=150$ pc for estimating the CV luminosity
function below. 

\section{Luminosity function}

We employed the $1/V_\mathrm{max}$ method \citep{s68}
to construct the hard X-ray luminosity function of CVs detected by the
INTEGRAL all-sky survey. 

To take into account the inhomogeneous distribution of
sources around the Sun, we weight the standard $\delta V_\mathrm{max}$
volume found for each small solid angle $\delta\Omega$ (at Galactic
latitude $b$) of the survey by the space density of sources integrated
over $\delta\Omega$ and over distance from 0 to $d_{\rm max}$, the
maximum distance at which a given source can be detected. This, for an
exponential distribution of sources, $\rho_{\rm CV}\propto
\exp(-z/h)$, leads to a generic volume \citep{trm93,sbb+02,sazonov06}
$$
\delta V_\mathrm{gen}=\delta\Omega\frac{h^3}
{\sin^3 |b|}\left[2-(\xi^2+2\xi+2)e^{-\xi}\right], 
$$
where $\xi=d_{\rm max}\sin |b|/h$. Each sampled source thus
contributes $1/\sum\delta V_\mathrm{gen}$ to the estimated 
space density and $1/(\sum\delta V_\mathrm{gen})^2$ to the associated variance,
where the sum is taken over the total solid angle of the survey. As
noted above, we adopt the scale height $h=150$ pc in our analysis.

\begin{figure}[htb]
\centerline{
\includegraphics[width=0.8\columnwidth,bb=18 170 567 700,clip]{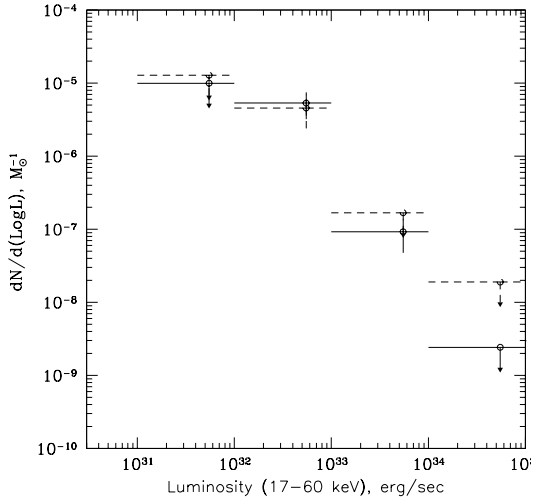}
}
\caption{Luminosity function of CVs detected by the INTEGRAL
all-sky survey (solid crosses). Dashed crosses show the luminosity
function constructed from those INTEGRAL CVs located at $|b|>5^\circ$, where
the survey identification is almost complete. Arrows denote  95\% upper
limits on the space density.
}
\label{lumfunc_int}
\end{figure}

The luminosity function of CVs detected by INTEGRAL is presented in
Fig.~\ref{lumfunc_int}. It was normalized to the local stellar density
$\rho_\ast=0.04 M_\odot$/pc$^{3}$ \citep{jw97}. We present 95\%
($\sim2\sigma$) upper limits on the space density of CVs in the
$10^{31}$--$10^{32}$ and $10^{34}$--$10^{35}$~erg~s$^{-1}$ luminosity
intervals, where we have 0 and 1 sources, respectively. 

The integrated space density of CVs with hard X-ray luminosity
$L_{\rm 17-60~keV}>10^{32}$ \lum\ is $\rho_{\rm 
CV}=(3.8\pm1.5)\times10^{-6}$ $M^{-1}_\odot$.

It is important to note here that at present the INTEGRAL all-sky survey
catalog \citep{krivonos07b} has significant incompleteness due to the
unknown nature of a number of sources: 25 persistent sources still
do not have secure identifications. However, most of these sources
reside very close to the Galactic plane and many of them are likely high-mass
X-ray binaries, rather than CVs considered here.  

\begin{figure}[htb]
\centerline{
\includegraphics[width=0.8\columnwidth,bb=56 180 563 540,clip]{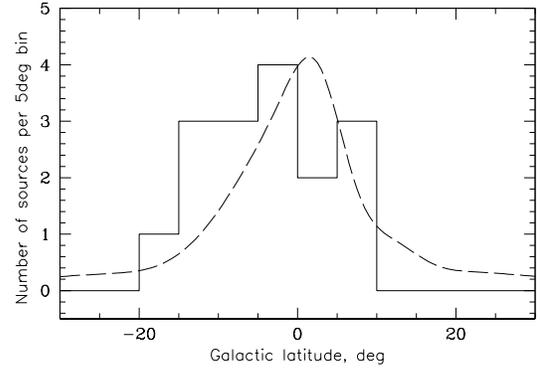}
}
\caption{Comparison of the Galactic latitude distribution of CVs
observed by INTEGRAL (histogram) with that expected for an
exponential distribution of CV space density, $\rho_{\rm CV}\propto
\exp(-z/150\, {\rm pc})$, taking into account the sensitivity map of
the survey.  
}
\label{distrib}
\end{figure}

If there were significant incompleteness in our CV sample, it would
manifest itself in the observed distribution of CVs over the sky. We
therefore calculated the expected numbers of CVs in 5-deg wide
Galactic latitude bins assuming the exponential density distribution
$\rho_{\rm CV}\propto \exp(-z/150\, {\rm pc})$ and luminosity
function $N(>L)= 1.52\times10^{-7} (L_{\rm 17-60~keV}/10^{32} {\rm
erg~s}^{-1})^{-1.5}$ pc$^{-3}$ (at $L_{\rm 17-60~keV}>10^{32}$ \lum),
and taking into account the sensitivity map of the survey. Comparison of the
observed and expected distributions (Fig.~\ref{distrib}) suggests that
our CV sample is not strongly incomplete.

We can also make more quantitative assessment of possible sample
incompleteness on our results. In Fig.~\ref{lumfunc_int} we show the
luminosity function constructed from the CVs detected away from
the Galactic plane ($|b|>5^\circ$), where the INTEGRAL all-sky survey
catalog is highly complete (2 unidentified sources vs. 12 CVs). The
good agreement of this luminosity function with the all-sky one
indicates that our determination is robust. 

\subsection{Comparison with the luminosity function of CVs detected by
the RXTE all-sky survey}

\begin{figure}[htb]
\centerline{
\includegraphics[width=0.8\columnwidth,bb=54 180 570 700,clip]{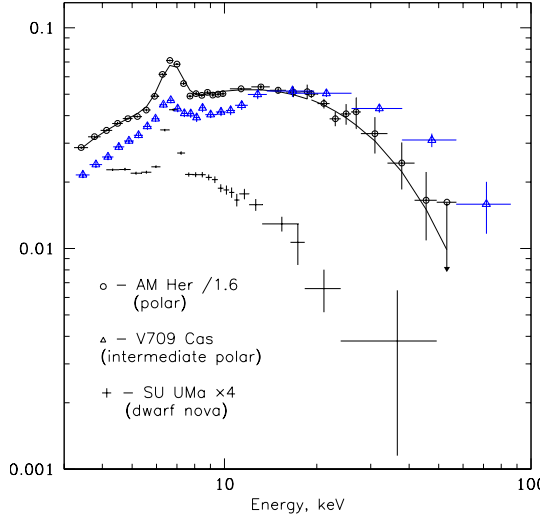}
}
\caption{Typical broadband spectra of different classes of
CVs, indicating that INTEGRAL observations in the hard X-ray energy
band (17--60 keV) are biased toward detecting the hardest CVs --
intermediate polars. 
}
\label{spectra_cvs}
\end{figure}

The previous measurement of the spatial distribution and X-ray luminosity 
function of accreting white dwarfs was done using the RXTE slew survey at
$|b|>10^\circ$ in the energy band 3--20 keV \citep{sazonov06}. We might expect 
that the harder energy band of the INTEGRAL survey should lead to
preferential selection of the hardest sources among all sub-classes of
CVs. Indeed, in our current INTEGRAL sample of sources the vast
majority are intermediate polars, which are known to be the hardest of all CVs
(see e.g. \citealt{sazonov06} and Fig.~\ref{spectra_cvs}).

The broadband (1--100~keV) spectra of IPs are typically so hard
(see e.g. \citealt{suleimanov05}) that the luminosities emitted in the
3--20 and 17--60 keV energy bands are similar. The ratio of
luminosities in these bands does have some scatter but is typically
$L_{\rm 17-60~keV}/L_{\rm 3-20~keV} \sim$0.6--0.9. Therefore, by dividing the
luminosities of IPs measured by INTEGRAL in the 17--60~keV by this
coefficient we can compare the CV luminosity function derived in this
work with that determined with RXTE in the 3--20 keV energy band
\citep{sazonov06}. 

\begin{figure}[htb]
\centerline{
\includegraphics[width=0.8\columnwidth,bb=18 170 567 700,clip]{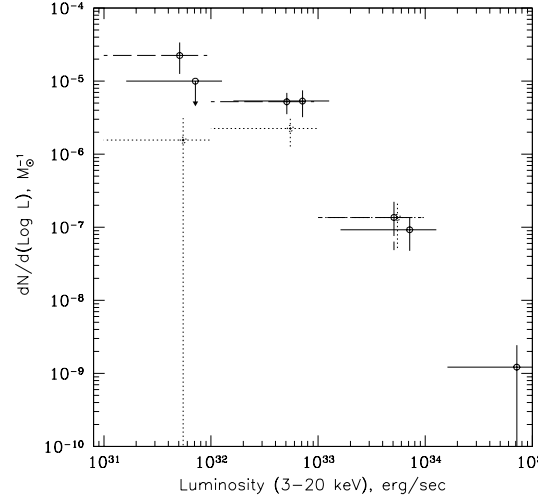}
}
\caption{Luminosity function of CVs (mostly intermediate polars)
detected by the INTEGRAL all-sky survey (solid crosses), converted to
the 3--20~keV energy band assuming the hardness ratio $L_{\rm 17-60 keV}/L_{\rm
3-20 keV}=0.7$ for IPs. This luminosity function is compared with the
luminosity function of faint Galactic sources detected by the RXTE
slew survey at $|b|>10^\circ$ (dashed crosses), with dotted crosses showing the
contribution of IPs to this luminosity function \citep{sazonov06}.
Note that, in contrast to Fig.~\ref{lumfunc_int}, we show here a rough
estimate of the space density in the $L_{\rm
17-60~keV}=10^{34}$--$10^{35}$~\lum\ interval based on the single
source detected by INTEGRAL in this range, rather than an upper limit.
}
\label{lumfunc}
\end{figure}

The result of this comparison is shown in Fig.~\ref{lumfunc}. There is
good agreement between the INTEGRAL and RXTE luminosity
functions at luminosities $L_{\rm 3-20~keV}>10^{32}$~erg~s$^{-1}$, 
where most of CVs are intermediate polars. At lower luminosities,
other subclasses of CVs become numerous, but due to their relatively
soft spectra such sources are practically absent in the INTEGRAL (hard X-ray
selected) sample. 

\section{Hard X-ray luminosity density of CVs}

\begin{figure}[htb]
\includegraphics[width=\columnwidth]{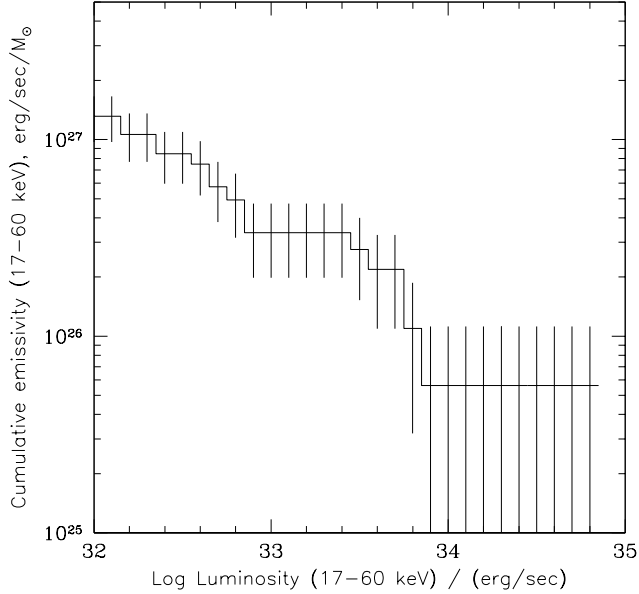}
\caption{Cumulative hard X-ray luminosity density of CVs determined
from the INTEGRAL all-sky survey.
}
\label{cumemiss}
\end{figure}

Using the same $1/V_\mathrm{max}$ method we can also estimate the
cumulative hard X-ray emissivity of CVs: 
$$
EM=\sum_i{L_i\over{\sum\delta V_\mathrm{max}(L_i)}},
$$
$$
\delta EM=\sqrt{\sum_i{\left(L_i\over{\sum\delta
V_\mathrm{max}(L_i)}\right)^2}}.
$$
We find that the cumulative 17--60 keV emissivity
of CVs with $L_{\rm 17-60~keV}>10^{32}$ \lum\ per unit stellar mass is
$L_{\rm 17-60~keV}/M_\ast=(1.3\pm0.3)\times10^{27}$ \lum\
$M^{-1}_\odot$. Fig.~\ref{cumemiss} shows the cumulative 
emissivity as a function of the threshold luminosity.

The obtained value of the CV cumulative emissivity is in remarkably
good agreement with the emissivity of the unresolved hard X-ray
emission observed by INTEGRAL from the Galactic plane region: 
$L_{\rm 17-60 keV}/M_\ast=(0.9-1.2)\times10^{27}$ \lum\ $M^{-1}_\odot$
\citep{krivonos07a}, suggesting that this emission is mostly the
superposition of numerous cataclysmic variables. 

\section{Broadband hard X-ray luminosity function of the old stellar
population in the Milky Way}

\begin{figure}[htb]
\includegraphics[width=\columnwidth]{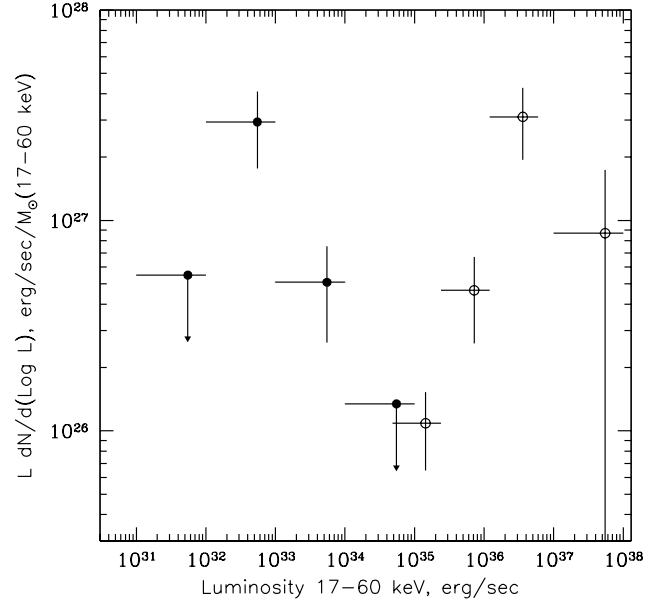}
\caption{Combined luminosity function of CVs ($L_{\rm
hx}<10^{34.5}$~erg~s$^{-1}$, filled circles) and LMXBs ($L_{\rm
hx}>10^{34.5}$ erg~s$^{-1}$, open circles).
}
\label{lumfunc_wide}
\end{figure}

It has been known since early observations with focusing X-ray telescopes that 
the cumulative emission of bright ($L_{\rm x}>10^{37}-10^{39}$ \lum)  
discrete sources constitutes a significant fraction of the total X-ray
emission of galaxies \citep{tf85}. This is especially true for energies
$>$2--3~keV, at which emission from hot interstellar gas cannot contribite
significantly even in gas rich galaxies because typical gas temperatures
are $kT\la1$ keV. Therefore, it is often assumed that most of the hard X-ray
emission of galaxies is provided by low- and high-mass X-ray binaries,
while the contribution of weaker sources is usually neglected. 

However, as we have shown in the previous section, CVs are expected to
produce a hard X-ray emissivity comparable to that generated by
brighter sources -- low-mass X-ray binaries. The same was previously 
suggested by \cite{krivonos07a} based on a study of the Galactic bulge
with INTEGRAL in which the luminosity of the unresolved hard X-ray emission was
found to be comparable to the cumulative luminosity of bright resolved
sources. We therefore construct below the combined hard X-ray luminosity
function of CVs and LMXBs.

In Fig.~\ref{lumfunc_wide} we present the luminosity function of CVs
and LMXBs in the 17--60~keV energy band based on INTEGRAL data
analysed in the present paper and in \cite{revnivtsev08}. This plot
demonstrates that the cumulative hard X-ray luminosity density of
LMXBs is only marginally larger than that of CVs. 
 
This ratio of the cumulative emissivities of LMXBs and CVs in the hard
X-ray band is significantly smaller than in the standard X-ray  
band (2--10~keV) see e.g. \citealt{sazonov06}). The major reason for
this is that the brightest LMXBs ($L_{\rm x}\sim10^{38}$ \lum\ in the
2--10~keV band) which provide the majority of cumulative
emissivity of LMXBs have very soft spectra and thus emit only a small
fraction of their total emission in hard X-rays. At the same time,
luminous CVs emit almost the same fraction of their bolometric
luminosity in standard and hard X-rays.

\section{Conclusions}

We have studied a sample CVs detected in the hard X-ray band
(17--60~keV) by the INTEGRAL all-sky survey. This sample is one of the
largest unbiased samples of luminous intermediate polars existing today.

Our findings are:
\begin{enumerate}

\item The scale height of the Galactic disk distribution of CVs is
$h=130^{+93}_{-46}$ pc. 

\item We have measured the luminosity function of CVs in the 17--60 keV energy 
band. The space density of CVs with luminosities $L_{\rm 17-60~keV}>10^{32}$ 
\lum\ is $(3.8\pm1.5)\times10^{-6}  M^{-1}_\odot$. 

\item The luminosity density of CVs with luminosities $L_{\rm
17-60~keV}>10^{32}$ \lum\ is $L_{\rm   
17-60~keV}/M_\ast=(1.3\pm0.3)\times10^{27}$ \lum\ $M^{-1}_\odot$. This
value is in remarkably good agreement with the emissivity of the
unresolved (with INTEGRAL) hard X-ray emission in the Galactic plane region.
\end{enumerate}

We emphasize that all our estimates depend on the
accuracy of the distance determination for our sources. 
Therefore, the current lack of accurate distances for a number of
sources from the sample introduces additional (not included
in the errors quoted above) uncertainties in the measured parameters of the
Galactic population of cataclysmic variables. 

\begin{acknowledgements}
This research made use of data obtained from the High Energy Astrophysics
Science Archive Research Center Online Service, provided by the
NASA/Goddard Space Flight Center.
This work was supported by DFG-Schwerpunktprogramme (SPP 1177)

\end{acknowledgements}

\end{document}